\documentclass[12pt]{iopart}
\usepackage{iopams}
\usepackage{epsfig}   
\usepackage{graphics}
\usepackage{url}
\newcommand{\apj}{Astrophys. J.}

\newcommand{\apjs}{Astrophys. J. Suppl.}
\newcommand{\aj}{Astron. J.}
\newcommand{\aap}{Astron. Astrophys.}
\newcommand{\prd}{Phys. Rev. D}
\newcommand{\physrep}{Phys. Rept}
\newcommand{\mnras}{Mon. Not. Roy. Astron. Soc.}
\begin{document} 

\title[Constraining dark energy fluctuations with supernova correlations]
{Constraining dark energy fluctuations with supernova correlations}

\author{Michael Blomqvist$^{1}$, Jonas Enander$^{2}$ and Edvard M\"ortsell$^{2}$}

\address{$^1$ The Oskar Klein Centre for Cosmoparticle Physics, Department of Astronomy, \\
Stockholm University, AlbaNova University Center \\
SE--106 91 Stockholm, Sweden}
\address{$^2$ The Oskar Klein Centre for Cosmoparticle Physics, Department of Physics, \\
 Stockholm University, AlbaNova University Center \\
SE--106 91 Stockholm, Sweden}
\ead{\mailto{michaelb@astro.su.se}, \mailto{enander@fysik.su.se}, \mailto{edvard@fysik.su.se}}

\begin{abstract}
We investigate constraints on dark energy fluctuations using type Ia supernovae. 
If dark energy is not in the form of a cosmological constant, that is if the equation of state $w\neq-1$, we expect not only temporal, but also spatial variations in the energy density. Such fluctuations would cause local variations in the universal expansion rate and  directional dependences in the redshift-distance relation. We present a scheme for relating a power spectrum of dark energy fluctuations to an angular covariance function of standard candle magnitude fluctuations.
The predictions for a phenomenological model of dark energy fluctuations   
are compared to observational data in the form of the measured angular covariance 
of Hubble diagram magnitude residuals for type Ia supernovae 
in the Union2 compilation. The observational result is consistent with 
zero dark energy fluctuations. However, due to the limitations in statistics, 
current data still allow for quite general dark energy fluctuations as long as 
they are in the linear regime.
\end{abstract}


\section{Introduction}
The current standard model of cosmology has the following, at least approximate, properties~\cite{2010ApJ...716..712A,2010arXiv1001.4538K,2010MNRAS.401.2148P}:
\begin{itemize}
	\item It is spatially flat, i.e., the total density is close to the critical density.
	\item 5\,\% of this density is made up of baryonic matter.
	\item 25\,\% is made up of pressureless dark matter, so far only observed through its gravitational effect on the universal expansion, structure formation and the dynamics of collapsed structures down to the scale of dwarf galaxies.
	\item 70\,\% in a cosmological constant, observed through an apparent accelerated universal expansion at low redshifts.
\end{itemize}
However, there is still room for sizeable deviations from the concordance picture. For example, current data allow for the dominant energy contribution to have properties quite different from that of a cosmological constant. 
Although the equation of state, relating the pressure and the density by a linear relation, $p=w\rho$, is probably close to the cosmological constant value $w=-1$ at low redshifts, it may have very different properties at higher redshifts~\cite{2010ApJ...716..712A}.

A dark energy component with $w\neq -1$ not only allows for a time varying energy density, but will also exhibit spatial fluctuations. In recent years, several studies have been carried out concerning clustering properties of different dark energy models (see, e.g., Mota~\etal~\cite{2008ApJ...675...29M} and references therein). Depending on the dark energy model employed, clustering can occur on very different scales and at different amplitudes and have different couplings to the dark matter clustering. As a concrete example of clustering scales and amplitudes in the non-linear regime, it was shown in Mota~\etal~\cite{2008ApJ...675...29M} that the density contrast of a dark energy scalar field can be of order $10^{-3}\left(1+w\right)$ in voids of radii $100-300$~Mpc. The clustering can be anti-correlated with the dark matter clustering, which is the case for, e.g., a quintessence field with a quadratic potential in the linear regime~\cite{2007PhRvD..75f3507D}. Clustering of dark energy may also affect structure growth in the dark matter component by, e.g., introducing modifications to the spherical collapse model~\cite{2004A&A...421...71M}. In short, there are no definite theoretical predictions on how dark energy will cluster in the universe. Coupled to the fact that it is notoriously difficult to measure temporal variations in the dark energy density, this warrants observational constraints on the amount of spatial clustering of dark energy in order to rule out or to further confirm the cosmological constant hypothesis.

Dark energy clustering will manifest itself through several different observational effects. Dark energy inhomogeneities will change the form and time evolution of gravitational potentials to affect cosmic microwave background (CMB) anisotropies and the late-time integrated Sachs-Wolfe (ISW) effect,  structure growth will be modified, and the properties of collapsed halos can change to affect cluster number counts~\cite{2003MNRAS.346..987W, 1998PhRvL..80.1582C, 2006PhRvD..73h3502K, 2010CQGra..27j5011G, 2010PhRvD..81j3513D, 2004PhRvD..70l3002H, 2009PhRvD..79b3502D,2004PhRvD..69h3503B,2005PhRvD..71l3521C,2006PhRvD..74d3505T,2006MNRAS.371.1373M, 2006A&A...450..899N, 2009JCAP...07..040A}.

In this paper we consider how dark energy clustering can affect observations of type Ia supernovae (SNe~Ia). While the ISW effect takes into account the impact local inhomogeneities have on photons arriving from the last scattering surface at redshift $z\approx1090$, observations of SNe~Ia probe light propagation out to $z\approx1.4$. 
When light from a SN~Ia passes through a region where the dark energy density fluctuates, the inferred luminosity distance will change due to the local expansion perturbation that the fluctuation creates. This will show up as correlated fluctuations in the SN~Ia peak magnitudes, where the magnitude correlation amplitudes and angular scales will depend on the specific dark energy clustering amplitudes and scales. A framework for constraining dark energy clustering models with SN~Ia data was considered in Cooray~\etal~\cite{2008arXiv0812.0376C}. Observational constraints on the amount of correlation in the magnitude fluctuations were established in Blomqvist~\etal~\cite{2008JCAP...06..027B}. Directional dependence in the SN data can also be used to probe models with an anisotropic dark energy equation of state~\cite{2008ApJ...679....1K,2008JCAP...06..018K} or to search for deviations from an isotropic expansion rate~\cite{2001MNRAS.323..859K,2007A&A...474..717S,2010MNRAS.401.1409C,2010arXiv1005.2868G}. An anisotropic relation between the redshift and magnitude of SNe~Ia is also expected in models where an observer is located off-center in a spherically symmetric void in the matter distribution~\cite{2010JCAP...05..006B,2007PhRvD..75b3506A}.

Detecting correlations in SN~Ia data is demanding because of the scatter in the peak magnitudes, arising both from intrinsic variation and from systematic effects that could induce correlated fluctuations. The observed peak magnitude is a derived quantity that depends sensitively on the details of how the SN~Ia luminosities are standardised~\cite{2009ApJS..185...32K}. Also, a number of physical effects may contaminate the results. At low redshift, 
correlated fluctuations can arise from large-scale peculiar motions~\cite{2007ApJ...661..650H,2006PhRvD..73j3002C}. At higher redshift and small angular separations, the dominating physical effect is gravitational lensing~\cite{2006PhRvL..96b1301C}. Intervening galactic dust may also induce correlations on small angular scales~\cite{2007ApJ...657...71Z}.

In this paper, we consider the framework presented in Cooray~\etal~\cite{2008arXiv0812.0376C} and continue the development of the necessary formalism for constraining dark energy spatial fluctuations. We also point out some of the limitations that the formalism carries.

The paper is organised as follows: In Section~\ref{data}, we describe the method to analyse SN~Ia magnitude residuals for angular correlations and apply it to current data to establish observational constraints on the amount of correlation. In Section~\ref{fluctuations}, we develop a theoretical formalism to relate a phenomenological power spectrum of dark energy spatial fluctuations to an angular covariance in the magnitude fluctuations. Our main results are presented in Section~\ref{constraints} where we compare the angular covariance predicted from different models to the observational constraints. This paper is concluded in Section~\ref{conclusions}.

\section{Angular correlations in supernova data}\label{data}

In this section, we investigate whether there are correlated fluctuations in the observed peak magnitudes of SNe~Ia at different angular separations. We first give a general description of how we calculate the angular covariance function and then determine the observational constraints on the amount of correlation over the entire sky using current SN~Ia data. Throughout the paper we will assume a flat universe.

\subsection{Method}
The luminosity distance to a SN~Ia at redshift $z$ in a spatially flat, homogeneous universe is given by
\begin{equation}\label{eq:lumdist2}
d_{\rm L}=\frac{c\left(1+z\right)}{H_{0}}\intop_{0}^{z}\frac{dz^{\prime}}{E\left(z^{\prime}\right)}\ ,
\end{equation}
where
\begin{equation}\label{eq:hubblefunction1}
E\left(z\right)\equiv\frac{H(z)}{H_{0}}=\sqrt{\Omega_{\rm m}\left(1+z\right)^{3}+\Omega_{\rm x}f\left(z\right)}\ .
\end{equation}
Here we have assumed a generic dark energy component with an equation of state $w_{\rm x}\left(z\right)$ whose energy density evolves with redshift as
\begin{equation}\label{eq:deff(z)}
\rho_{\rm x}\left(z\right)=\rho_{{\rm x},0}f\left(z\right)=\rho_{{\rm x},0}\exp\left[ 3\intop_{0}^{z}dz^{\prime}\frac{1+w_{\rm x}\left(z^{\prime}\right)}{1+z^{\prime}}\right]\ .
\end{equation}
The apparent magnitude $m$ of the SN~Ia is related to the luminosity distance as
\begin{equation}
m=5\log _{10}\left(\frac{d_{\rm L}}{1\ \rm Mpc}\right)+M+25\ ,
\end{equation}
where $M$ is the absolute magnitude.

We apply a methodology similar to that outlined in Blomqvist~\etal~\cite{2008JCAP...06..027B} to investigate angular correlations in SN~Ia data. For all SNe~Ia we have redshifts, $z$, observed peak magnitudes, $m_{\rm obs}$, with uncertainties, $\sigma_{m}$, as well as positions in the sky. We form all possible SN~Ia pairs and calculate their angular separations, $\theta$, in the sky.

The magnitude residual, $\delta m$, of a SN~Ia on the Hubble diagram is the difference between its observed peak magnitude and the magnitude predicted from its redshift in the best fit cosmology, $m_{\rm fit}$,
\begin{equation}\label{eq:magres}
\delta m=m_{\rm obs}-m_{\rm fit}\ .
\end{equation}

The angular (auto)covariance function as a function of angular separation is defined as
\begin{equation}\label{covfunc}
cov(\theta)=\langle (\delta m-\mu_{\delta m})\cdot (\delta m(\theta)-\mu_{\delta m})\rangle\ .
\end{equation}
The covariance function gives a statistical measure of the amount of correlation of the magnitude residuals at different angular separations. If $cov(\theta)>0$, then SNe~Ia separated by the angle $\theta$ on average have correlated magnitude residuals, if $cov(\theta)<0$, the magnitude residuals are anti-correlated. Uncorrelated magnitude residuals will give $cov(\theta)=0$. Since the uncertainties $\sigma_{m}$ vary between SNe~Ia, we weigh each magnitude residual using the weights $1/\sigma_{m}^2$ in the calculations. $\mu_{\delta m}$ is thus the weighted mean of all the magnitude residuals and the outer brackets in equation~(\ref{covfunc}) denote a weighted expectation value. Writing $\delta m^{\prime}=\delta m-\mu_{\delta m}$, we have explicitly that
\begin{equation}
cov(\theta)=\left(\sum_{i,j}\frac{\delta m_{i}^{\prime}\delta m_{j}^{\prime}\left(\theta\right)}{\sigma_{m,i}^{2}\sigma_{m,j}^{2}}\right)\cdot\left(\sum_{i,j}\frac{1}{\sigma_{m,i}^{2}\sigma_{m,j}^{2}}\right)^{-1},
\end{equation}
where the summation is over all SN pairs with angular separation $\theta$. Using this procedure, we expect the dispersion in the covariance based on independent SN pairs with a given angular separation to be
\begin{equation}\label{eq:covdisp1}
\sigma_{cov}=\left(\sum_{i,j}\frac{1}{\sigma_{m,i}^{2}\sigma_{m,j}^{2}}\right)^{-\frac{1}{2}}\ .
\end{equation}
For an ideal set of SNe~Ia with equal uncertainties $\sigma_m$, the dispersion would reduce to
\begin{equation}
\sigma_{cov}=\frac{\sigma_{m}^2}{\sqrt{N_{\rm p}}}\ ,
\end{equation}
where $N_{\rm p}$ is the number of unique pairs with the given angular separation. Since each SN is included in many pairs, the actual dispersion will be slightly less than that given in equation~(\ref{eq:covdisp1}).

\subsection{Data set}
\begin{figure}
\begin{center}
\includegraphics[angle=0,width=.60\textwidth]{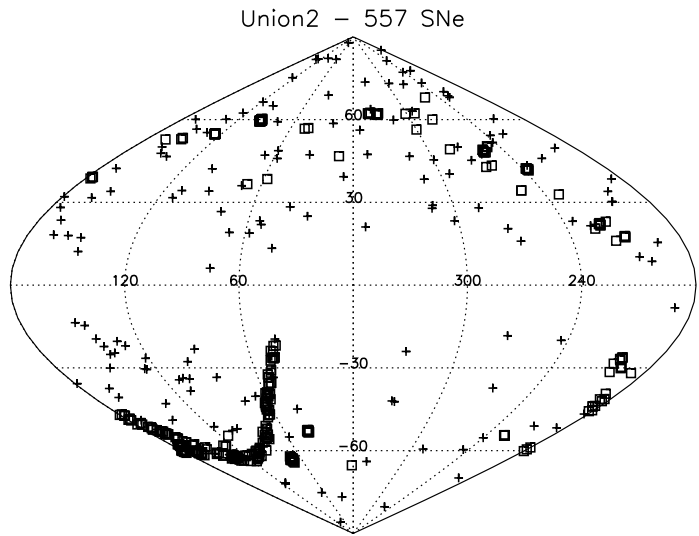}
\caption{\label{fig:union2map} Sky distribution in galactic coordinates of the 557 SNe~Ia in the Union2 data set. SNe~Ia with $z<0.1$ are marked with pluses and $z>0.1$ with squares. The distinct feature in the lower left part consists of SNe~Ia along the equatorial plane discovered by the SDSS-II survey.}
\end{center}
\end{figure}

For the data analysis, we employ the Union2 SN~Ia data set presented in Amanullah~\etal~\cite{2010ApJ...716..712A}. The compiled data set consists of 557 SNe~Ia covering the redshift range $z=[0.015,1.4]$ and augments the Union compilation~\cite{2008ApJ...686..749K} by adding SNe~Ia at low and intermediate $z$ discovered by the	CfA3~\cite{2009ApJ...700..331H} and SDSS-II Supernova Search~\cite{2008AJ....136.2306H}, respectively, as well as six new SNe~Ia discovered by the Hubble Space Telescope at high $z$. The peak magnitudes and uncertainties were obtained using the SALT2 light-curve fitter. The uncertainties include both the observational and the intrinsic magnitude scatter. Figure~\ref{fig:union2map} shows the sky distribution of the SNe~Ia in the Union2 data set. SNe~Ia with $z<0.1$ are marked with pluses and $z>0.1$ with squares. The sky distribution affects on what scales and to what level we can constrain the angular covariance function. Whereas the low $z$ SNe~Ia are distributed quite evenly across the sky, the SNe~Ia discovered at high $z$ are mainly confined to small survey patches. The most distinct feature in Figure~\ref{fig:union2map} is the dense collection of SNe~Ia along the equatorial plane discovered by the SDSS-II survey. We have augmented the Union2 data by adding the SN positions. The data used in the analysis are presented in Table~\ref{tab:data}\footnote{When used, please cite Amanullah~\etal~\cite{2010ApJ...716..712A} in addition to this paper. The Union2 data, including the covariance matrix with systematics, are available at \url{http://supernova.lbl.gov/Union}.}.

\begin{table}\footnotesize{
\begin{center}
\begin{tabular}[b]{lccccc}
\hline
SN & $z$ & $\mu$ & $\sigma_{\mu}$ & RA(J2000.0) & Dec(J2000.0)\\
\hline
1993ah & 0.0285 & 35.34 & 0.23 & 23:51:50.200 & -27:57:42.00\\
1993ag & 0.0500 & 36.68 & 0.17 & 10:03:34.500 & -35:27:45.00\\
1993o & 0.0529 & 36.82 & 0.16 & 13:31:08.370 & -33:12:54.60\\
1993b & 0.0701 & 37.44 & 0.16 & 10:34:52.180 & -34:26:30.20\\
1992bs & 0.0627 & 37.48 & 0.16 & 03:29:29.970 & -37:16:18.60\\
\hline
\end{tabular}
\caption{\label{tab:data}Union2 compilation. Redshifts are given in a frame at rest with respect to the CMB. The distance modulus is defined as $\mu=m-M$. This table is available in its entirety in a machine-readable form at \url{http://www.astro.su.se/~michaelb/SNdata}. A portion is shown here for guidance regarding its form and content.}
\end{center}
}
\end{table}

\subsection{Results}\label{snresults}
We fit a spatially flat model with dark energy with a constant equation of state and cold dark matter ($w$CDM model) to the data using the full covariance matrix with systematic errors. The best fit cosmological parameters are $w_{\rm x}=-1.2$ and $\Omega_{\rm m}=0.34$. The magnitude residuals are then obtained from equation~(\ref{eq:magres}).

\begin{figure}
\begin{center}
\includegraphics[angle=0,width=.60\textwidth]{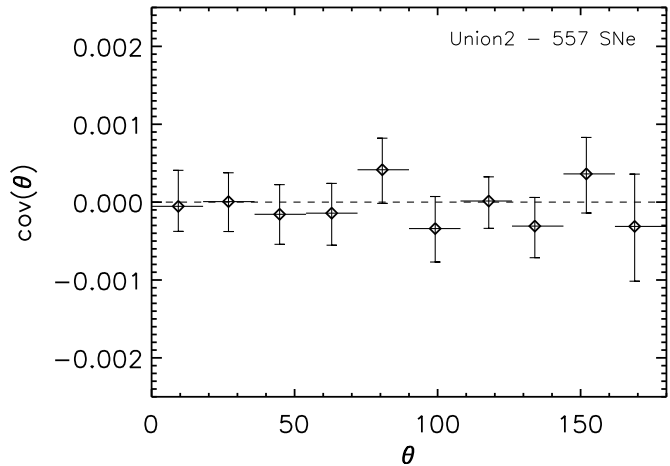}
\caption{\label{fig:sncorr_union2_cutz01} Angular covariance function for the 557 SNe~Ia in the Union2 data set. The error bars represent the 95\,\% confidence limit. The data have been binned uniformly using an angular resolution of $18^{\circ}$. The horizontal bars indicate the range of each bin and the points are placed at the
average angular separation in each bin. Note that the data 
points are not independent, since each SN contributes to several bins.}
\end{center}
\end{figure}

The angular covariance function for the 557 SNe~Ia in the Union2 data set is presented in Figure~\ref{fig:sncorr_union2_cutz01}. The data have been divided into ten uniform bins giving an angular resolution of $18^{\circ}$. Note that since each SN is included in several bins, the data points will not be independent. The error bars represent the 95\,\% confidence limit. The data set is consistent with the SN magnitude residuals being uncorrelated with the  covariance constrained to $<10^{-3}$ over the entire sky. The error bars are obtained using a bootstrap resampling method where we generate new data sets by keeping the positions in the sky fixed and randomly picking one of the SNe~Ia for each position. The covariance function is then calculated for 10~000 data sets to estimate the spread in the covariance function. We point out that our results are not particularly sensitive to what cosmology we subtract when calculating the magnitude residuals; using instead the best fit flat $\Lambda$CDM model has very small impact on the covariance function. In the next section, we investigate how we can use our observed limit on the angular correlation of SN~Ia magnitude residuals to constrain dark energy fluctuations. 

\section{Angular correlations from dark energy fluctuations}\label{fluctuations}

In this section, we consider how spatial fluctuations in the dark energy density induce correlated fluctuations in the SN~Ia magnitudes. We will employ a formalism that has previously been used in galaxy counts correlations and weak gravitational lensing analysis, and adapt it to relate the spatial power spectrum of dark energy fluctuations to an angular covariance function of magnitude fluctuations. Our goal is to estimate the possible effects that dark energy fluctuations can have on the SN covariance function which can be tested against current data. The formalism that we develop extends that presented in Cooray~\etal~\cite{2008arXiv0812.0376C} but is not without restrictions. As such, we only consider it to be a toy model, serving as a first step towards a more complete formalism. In our setup we use a phenomenological dark energy power spectrum as input for our model, neglecting the observed matter power spectrum. The main limitation, however, is that it is currently not possible to calculate exact distances in general inhomogeneous cosmologies, and we will have to rely on simplified models for relating dark energy fluctuations to luminosity distance fluctuations as explained below.   

\subsection{Power spectrum projection}
Suppose that there is a dark energy density fluctuation field defined by
\begin{equation}
\delta_{\rm x}\left(\mathbf{\hat{n}},z\right)\equiv\frac{\rho_{\rm x}\left(\mathbf{\hat{n}},z\right)-\bar{\rho}_{\rm x}\left(z\right)}{\bar{\rho}_{\rm x}\left(z\right)}\ ,
\end{equation}
where the bar indicates spatially averaged quantities and $\mathbf{\hat{n}}$ denotes directional dependence.
We assume that fluctuations in the dark energy density will perturb the expansion rate locally. Consider a perturbed Hubble function with perturbations in both the dark energy and matter component,
\begin{equation}\label{eq:perthubblefunc}
H^{2}\left(\mathbf{\hat{n}},z\right)=\frac{8\pi G}{3}\left[\bar{\rho}_{\rm m}\left(z\right)+\bar{\rho}_{\rm x}\left(z\right)+\delta\rho_{\rm x}\left(\mathbf{\hat{n}},z\right)+\delta\rho_{\rm m}\left(\mathbf{\hat{n}},z\right)\right]\ .
\end{equation}
The matter perturbation is introduced to balance the dark energy perturbation such that we recover $H\left(z=0\right)=H_{0}$, i.e., the Hubble constant does not have spatial variations. Alternatively, one could choose to normalise the local expansion such that the Hubble parameter is spatially homogeneous at an earlier time slicing, or to impose a condition of homogeneous Big Bang time for each local patch. Using this setup we can write (see \ref{append} for more details)
\begin{equation}
H^{2}\left(\mathbf{\hat{n}},z\right)=H_{0}^{2}\left[\Omega_{\rm m}\left(1+z\right)^{3}+\Omega_{\rm x}f\left(z\right)+\delta_{\rm x}\left(\mathbf{\hat{n}},z\right)\Omega_{\rm x}f\left(z\right)\alpha\left(z\right)\right]\ ,
\end{equation}
where
\begin{equation}\label{eq:alpha}
\alpha\left(z\right)=1-\frac{\left(1+z\right)^{3}}{f\left(z\right)}\left(1+z\right)^{-s_{\rm m}/2+s_{\rm x}/2}
\end{equation}
ensures that we have a spatially homogeneous expansion at $z=0$. The parameters $s_{\rm m}$ and $s_{\rm x}$ determine the degree of gravitational growth with redshift of the matter and dark energy density contrast, respectively. We point out that we have assumed a flat universe, and as a consequence the perturbations of dark energy and matter are necessarily anti-correlated. A less restricted setup of the perturbations would also allow for spatially varying curvature, but unfortunately it is currently not possible to calculate exact distances in such a general model.

The presence of dark energy inhomogeneities along the line of sight to a SN~Ia will cause fluctuations in the observed peak magnitude, or equivalently, a fractional fluctuation in the luminosity distance, related by
\begin{equation}\label{eq:apparentmagnitudedisturbance2}
\delta m=\frac{5}{\ln10}\frac{\delta d_{\rm L}}{d_{\rm L}}\ .
\end{equation}
The averaged fractional luminosity distance fluctuation integrated over all sources in the direction $\mathbf{\hat{n}}$, out to the most distant SN~Ia at comoving distance $r_{\rm s}$, is given by a projection of the dark energy fluctuation field down on the celestial sphere,
\begin{equation}\label{eq:averagefractionallumdist4}
\left[\frac{\delta d_{\rm L}}{d_{\rm L}}\left(\mathbf{\hat{n}}\right)\right]_{\rm average}=\intop_{0}^{r_{\rm s}}drW(r)\delta_{\rm x}\left[\mathbf{\hat{n}},z\left(r\right)\right]\ .
\end{equation}
The exact nature of the projection is determined by the weight function $W\left(r\right)$, derived in \ref{append},
\begin{equation}\label{eq:weightfunction}
W\left(r\right)=-\frac{\Omega_{\rm x}}{2E^{2}\left[z\left(r\right)\right]}f\left[z\left(r\right)\right]\alpha\left[z\left(r\right)\right]\intop_{r}^{r_{\rm s}}dr^{\prime}\frac{n\left(r^{\prime}\right)}{r^{\prime}}\ .
\end{equation}
The weight function basically tells us how much a dark energy fluctuation at a comoving distance $r$ will affect the luminosity distance of SNe at larger $r$. A key assumption in the derivation of the weight function is that the density fluctuations are small, i.e., dark energy clustering only occurs in the linear regime. Here $n\left(r\right)$ is the radial distribution of SNe~Ia, normalised so that
\begin{equation}\label{eq:normalizationcondition}
\intop_{0}^{r_{\rm s}}drn\left(r\right)=1\ .
\end{equation}

To calculate the induced correlations in SN~Ia magnitude fluctuations it is enough to know the statistical properties of the fluctuation field. If the dark energy density contrast is a Gaussian random field, these are encoded in the spatial power spectrum $P_{\rm x}\left(k\right)$, formally defined as
\begin{equation}
\left\langle \delta_{\rm x}\left(\mathbf{k}\right)\delta_{\rm x}^{*}\left(\mathbf{k}^{\prime}\right)\right\rangle =\left(2\pi\right)^{3}\delta_{\rm D}^{3}\left(\mathbf{k}-\mathbf{k}^{\prime}\right)P_{\rm x}\left(k\right)\ .
\end{equation}
Here $k$ is the comoving wavenumber related to the comoving wavelength $\lambda$ by $k=2\pi/\lambda$. The spatial power spectrum of dark energy fluctuations can be related to an angular power spectrum of fractional luminosity distance fluctuations $C_{l}$ using a projection in the flat-sky approximation involving the same weight function,
\begin{equation}\label{eq:fullformulapgpdelta}
C_{l}=\intop_{0}^{r_{\rm s}}dr\frac{W^{2}\left(r\right)}{r^{2}}P_{\rm x}\left(k=\frac{l}{r},z\left(r\right)\right)\ .
\end{equation}
The angular power spectrum tells us how much fluctuations there are on angular scales $\theta=180^{\circ}/l$. Equation~(\ref{eq:fullformulapgpdelta}) is a projection down on a small patch of the sky that effectively can be considered flat, and is only valid for small angular separations. This form of the flat-sky angular power spectrum requires that the weight function $W\left(r\right)$ is fairly constant over the regions of interest and that the correlation lengths are small compared to the integration range. This relation has been used extensively in relation to galaxy counts~\cite{1973ApJ...185..413P} and weak lensing observations~\cite{1998ApJ...498...26K,2001PhR...340..291B}. We note that a more general projection down on the entire celestial sphere, valid for all angular separations, is indeed possible to perform, but is more cumbersome to work with. Since we expect the induced correlations to be largest at small angular separations, equation~(\ref{eq:fullformulapgpdelta}) is sufficient for this study.

From the angular power spectrum of fractional luminosity distance fluctuations, we calculate the angular covariance function of magnitude fluctuations as
\begin{equation}\label{eq:covtheta}
cov\left(\theta\right)=\left(\frac{5}{\ln10}\right)^{2}\intop_{0}^{\infty}\frac{dl}{2\pi}lC_{l}J_{0}\left(l\theta\right)\ ,
\end{equation}
where $J_{0}$ is the zeroth order Bessel function of the first kind, and the prefactor comes from equation~(\ref{eq:apparentmagnitudedisturbance2}).

In summary, we use equations~(\ref{eq:weightfunction}), (\ref{eq:fullformulapgpdelta}) and (\ref{eq:covtheta}) to relate the spatial power spectrum of dark energy fluctuations to the angular covariance function of magnitude fluctuations.

\subsection{Phenomenological power spectrum of dark energy fluctuations}
In order to apply our formalism to constrain dark energy clustering, we need to specify a power spectrum of dark energy fluctuations that we can use to compare the theoretical covariance from the projected power spectrum with the observational covariance from SN~Ia data. As mentioned in the introduction, dark energy clustering is very model dependent. Thus, from a theoretical perspective, it is not possible to predict a single specific power spectrum. We will therefore use a phenomenological power spectrum $P_{\rm x}(k,z)$ which covers the amplitude distribution of different wavenumbers, with a cut-off towards small fluctuation scales and a redshift dependent growth factor. The specific parameterisation was previously considered in Cooray~\etal~\cite{2008arXiv0812.0376C} and is basically a modified inflation power spectrum. It is given by
\begin{equation}\label{eq:phenomenologicalformula}
\Delta^{2}\left(k\right)\equiv\frac{k^{3}P_{\rm x}\left(k,z\right)}{2\pi^{2}}=\delta_{\rm xH}^{2}\left(\frac{ck}{H_{0}}\right)^{n_{\rm x}+3}e^{-k/k_{\rm c}}\left(1+z\right)^{-s_{\rm x}}\ .
\end{equation}
Here we have also introduced the dimensionless quantity $\Delta^{2}(k)$, which gives the contribution to the variance of the fluctuation field per logarithmic interval. The power spectrum has four free parameters which can be constrained by data: $\delta_{\rm xH}^{2}$ gives the fluctuation amplitude on horizon scales; $n_{\rm x}$ gives the power law distribution of the fluctuations; $k_{\rm c}$ is a cut-off wavenumber above which fluctuations are suppressed; and $s_{\rm x}$ is the growth parameter from equation~(\ref{eq:alpha}).

In the analysis, we fix the growth parameters to $s_{\rm m}=s_{\rm x}=2$ corresponding to a growth directly proportional to the scale factor (i.e. the density contrast decreases with redshift) as expected for linear matter fluctuations in a matter dominated universe. Since we only integrate out to $z=1.4$ in the power spectrum projection, the final results are not sensitive to this choice.

\begin{figure}
\begin{center}
\includegraphics[angle=0,width=.48\textwidth]{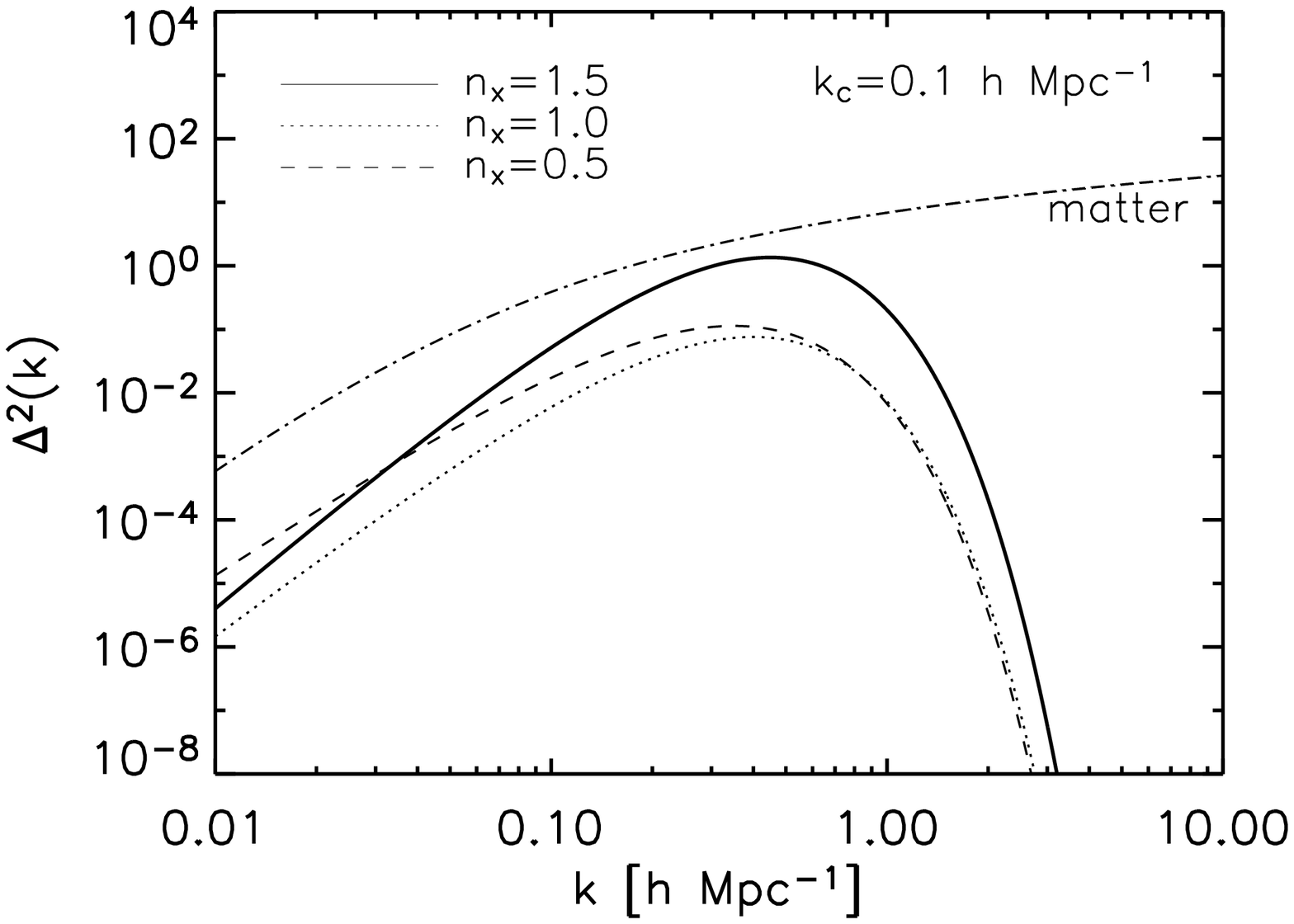}
\includegraphics[angle=0,width=.48\textwidth]{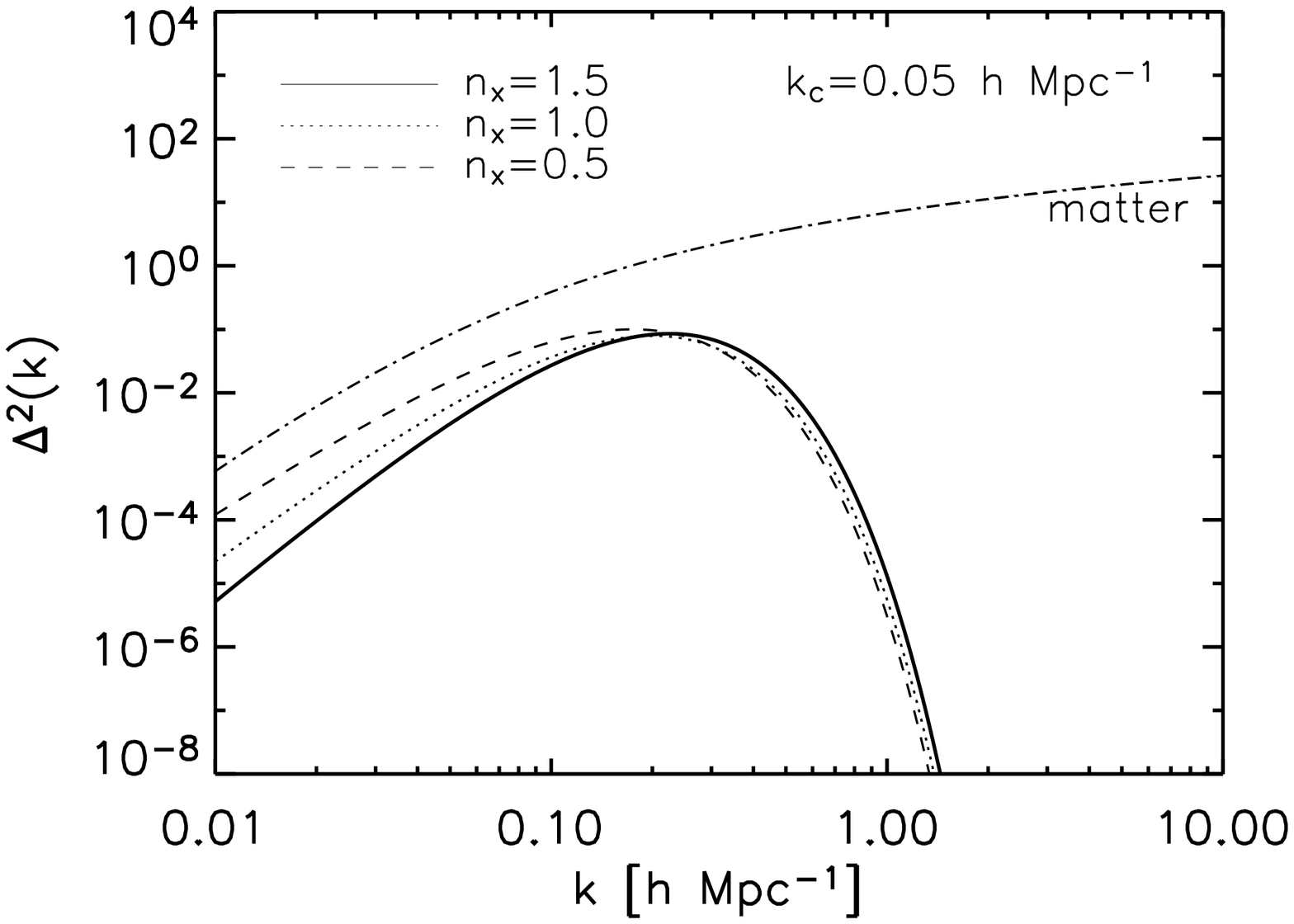}
\caption{\label{fig:delta_all} Phenomenological dark energy spatial power spectrum today for $n_{\rm x}=[1.5, 1, 0.5]$. The left panel shows models with cut-off wavenumber $k_{\rm c}=0.1$~$h$~Mpc$^{-1}$. The amplitude values are (in the above order) $\delta_{\rm xH}^{2}=[10^{-12},2\times10^{-12},10^{-10}]$. The right panel instead shows models with $k_{\rm c}=0.05$~$h$~Mpc$^{-1}$. Here the amplitudes are $\delta_{\rm xH}^{2}=[10^{-12},3\times10^{-11},10^{-9}]$. The dot-dashed line indicates the linear matter power spectrum from Peacock~\cite{1997MNRAS.284..885P}.}
\end{center}
\end{figure}

Figure~\ref{fig:delta_all} shows the power spectrum for different values of the parameters $n_{\rm x}$ and $k_{\rm c}$. The values considered ensure that the spectra tend to zero on large scales, where the projection formalism breaks down, and that fluctuations are suppressed on small scales where clustering is not expected. The curves have been normalised so that the variance of the fluctuation field is $\left\langle \delta_{\rm x}^{2}\right\rangle=0.1$, since we demand that the density fluctuations are small. This means that the amplitude $\delta_{\rm xH}^{2}$ has been adjusted for each individual model according to this normalization condition (see the figure caption for the different values used). Also shown for comparison is the linear matter power spectrum calculated using a fitting formula in Peacock~\cite{1997MNRAS.284..885P}. The dark energy power spectra all lie below the matter power spectrum. This need not necessarily be expected from a dark energy clustering model, but is instead only a consequence of imposing that the fluctuations are small in our formalism.

Given the power spectrum of dark energy fluctuations, we can calculate the angular power spectrum of fractional luminosity distance fluctuations using equation~(\ref{eq:fullformulapgpdelta}). Note that the angular power spectrum depends on the redshift distribution of SNe~Ia through the weight function. The radial SN distribution $n(r)$ is taken directly from the Union2 data set. The distribution is thus a sum of delta functions that pick out the specific distances to the different SNe~Ia such that the weight function $W\left(r\right)$ becomes
\begin{equation}\label{eq:weight2}
W\left(r\right)=-\frac{\Omega_{\rm x}}{2E^{2}\left[z\left(r\right)\right]}f\left[z\left(r\right)\right]\alpha\left[z\left(r\right)\right]\frac{1}{N_{\rm SN}}\sum_{r_{i}>r}{\frac{1}{r_{i}}}\ ,
\end{equation}
where $N_{\rm SN}$ is the number of SNe~Ia in the data set and the summation is over all distances larger than $r$ out to the most distant SN at $r_{\rm s}$. As the unperturbed background cosmology, we use the best fit flat $w$CDM model in Section~{\ref{snresults}}.

\begin{figure}
\begin{center}
\includegraphics[angle=0,width=.60\textwidth]{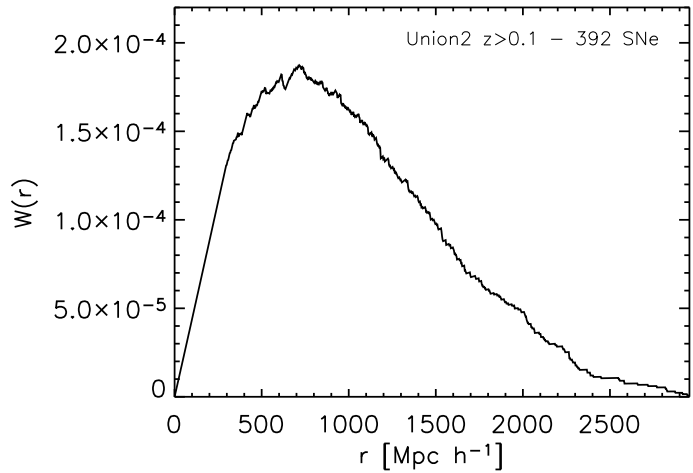}
\caption{\label{fig:weight_union2} Weight function used in the power spectrum projection. It is based on the radial distribution of the 392 SNe at $z>0.1$ in the Union2 data set.}
\end{center}
\end{figure}

We choose to make a redshift cut in the data set, excluding all SNe~Ia at $z<0.1$, for a couple of reasons. Firstly, the non-uniform radial distribution of SNe~Ia, which is particularly populated at low $z$, gives a spike in the weight function at $r=50-100$~$h^{-1}$~Mpc. The redshift cut removes this feature, giving a smoothly varying weight function as necessary in the power spectrum projection. Secondly, low $z$ SNe are affected by their peculiar motions which can induce correlated fluctuations in the observed peak magnitudes. Removing all SNe at $z<0.1$ is a conservative choice to obviate this effect. The redshift cut brings down the total number of SNe~Ia in the data set from 557 to 392 (see also Figure~\ref{fig:union2map} for which SNe that are removed). The observational limits used in the next section will essentially be unaffected by this cut. Figure~\ref{fig:weight_union2} shows the weight function in equation~(\ref{eq:weight2}) based on the radial distribution of the 392 SNe at $z>0.1$ in the Union2 data set. 

\begin{figure}
\begin{center}
\includegraphics[angle=0,width=.48\textwidth]{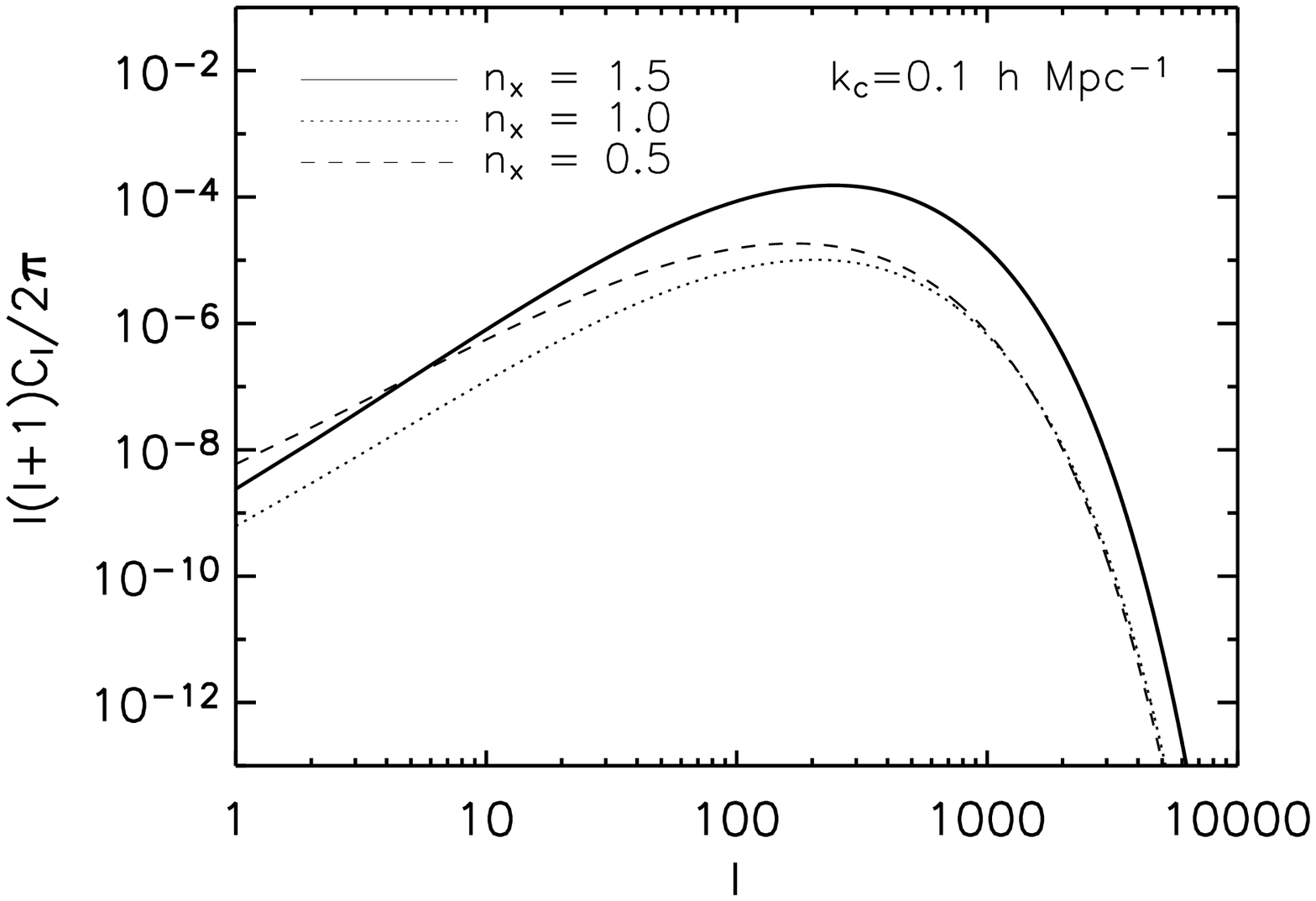}
\includegraphics[angle=0,width=.48\textwidth]{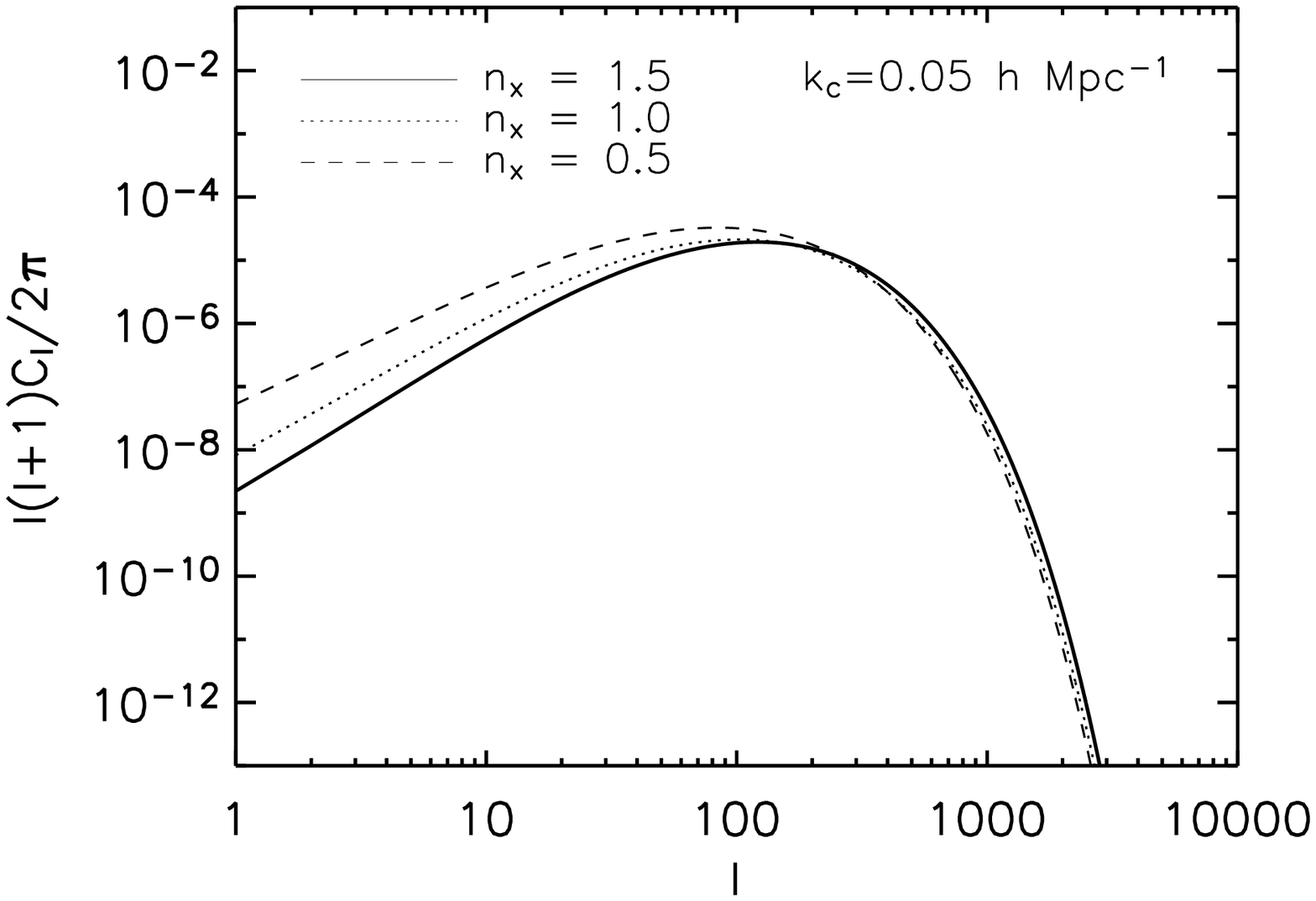}
\caption{\label{fig:lcl_all} Angular power spectra of fractional luminosity distance fluctuations for different values of the parameters $n_{\rm x}$ and $k_{\rm c}$. The curves are normalised as in Figure~\ref{fig:delta_all}.}
\end{center}
\end{figure}

Figure~\ref{fig:lcl_all} shows the angular power spectrum of fractional luminosity distance fluctuations. The amplitude of the curves are normalised according to the values in Figure~\ref{fig:delta_all}. It would have been interesting to compare this result to the angular power spectrum predicted in a model with perturbations in either dark energy or matter. However, the current formalism considers perturbations in both energy components and, as such, cannot address, e.g., luminosity distance fluctuations from matter perturbations in a $\Lambda$CDM universe.

\section{Constraints on dark energy spatial fluctuations}\label{constraints}

\begin{figure}
\begin{center}
\includegraphics[angle=0,width=.48\textwidth]{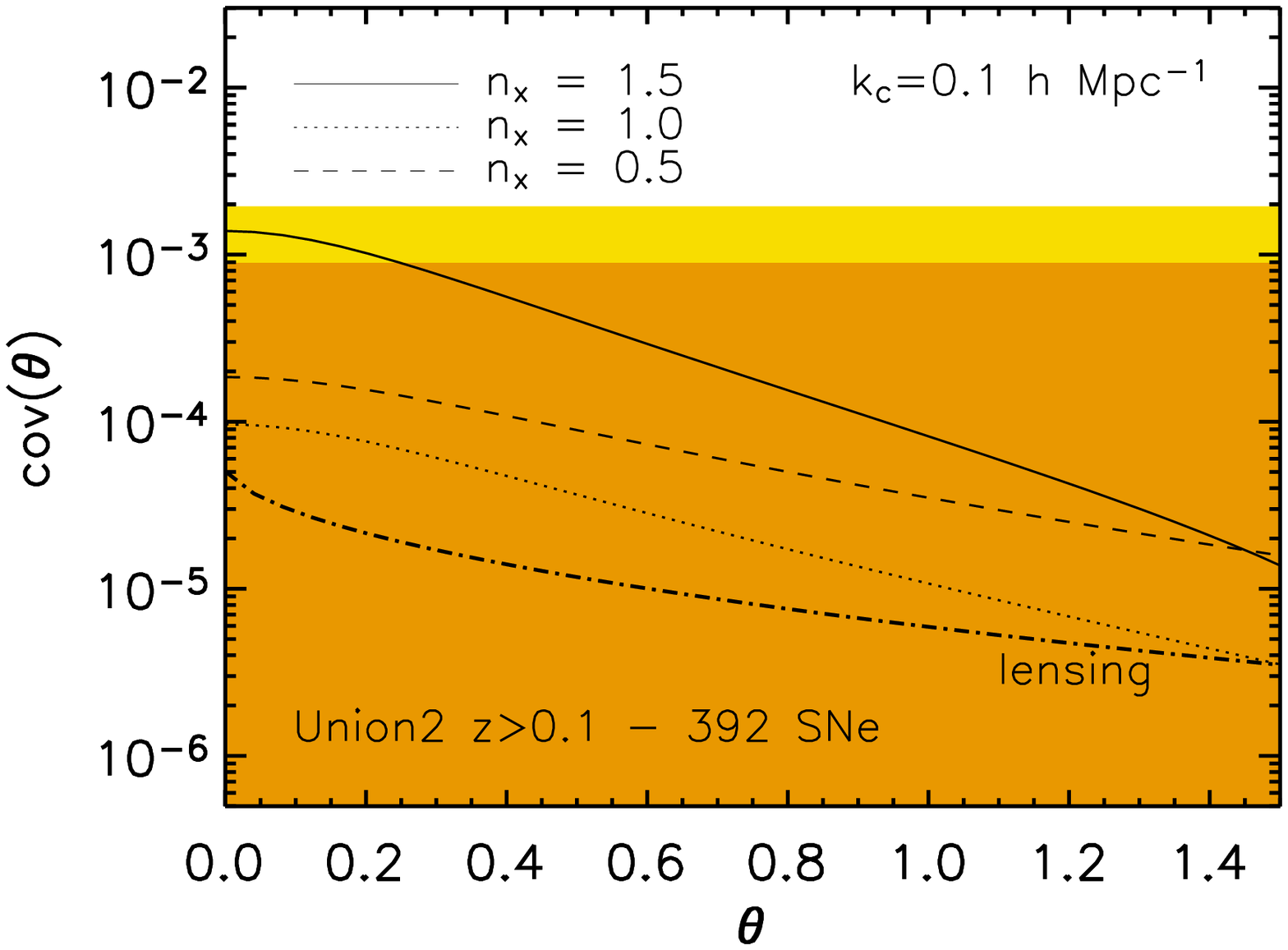}
\includegraphics[angle=0,width=.48\textwidth]{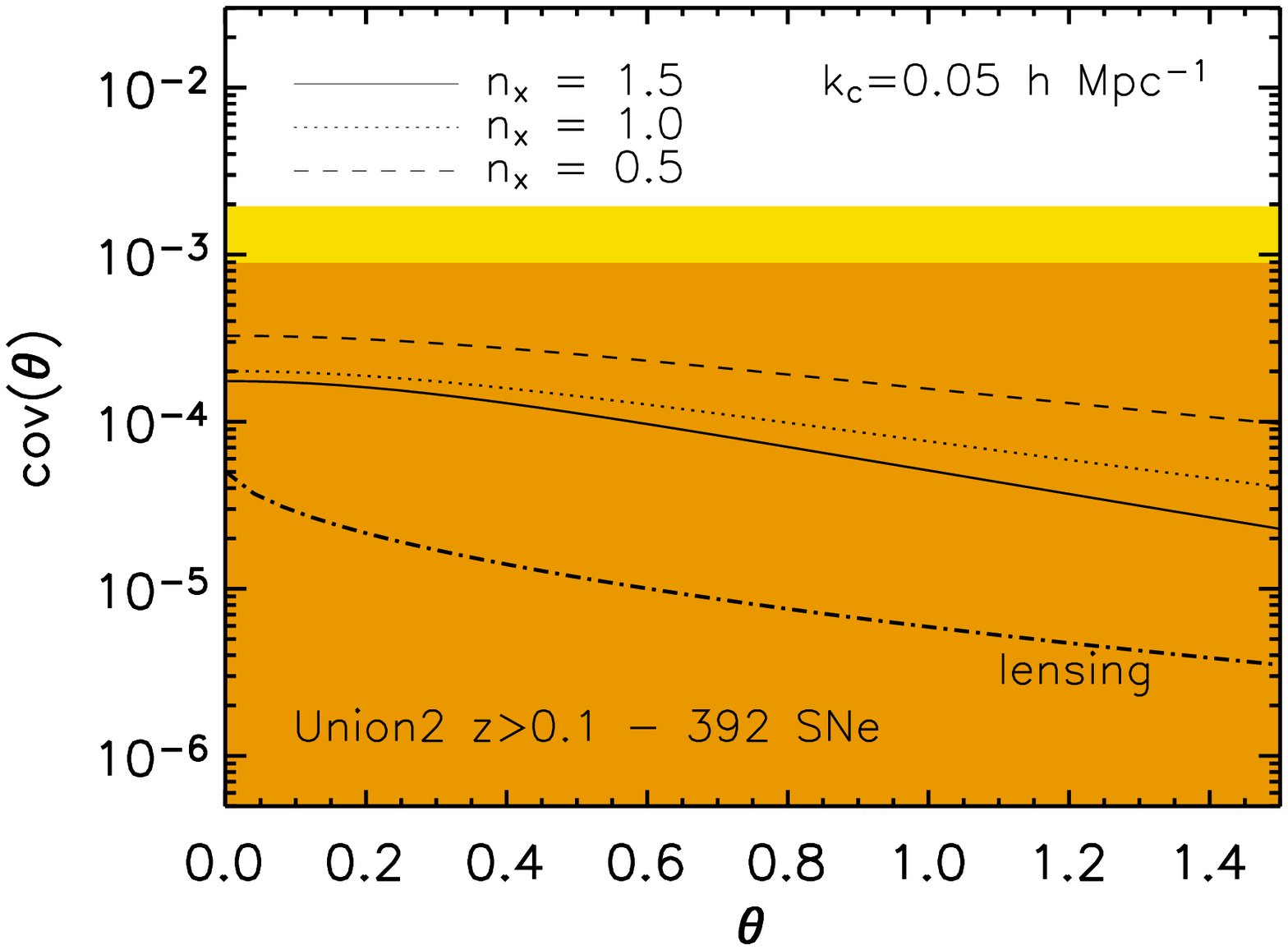}
\caption{\label{fig:lim_union2_nx} Angular covariance function of magnitude fluctuations for different values of the parameters $n_{\rm x}$ and $k_{\rm c}$ of the dark energy power spectrum. The curves are normalised as in Figure~\ref{fig:delta_all}. The confidence limits obtained from the data are 95\,\% (orange) and 99.7\,\% (yellow). Included is also the covariance expected from gravitational lensing (dot-dashed).}
\end{center}
\end{figure}

Using equation~(\ref{eq:covtheta}), we calculate the angular covariance function of magnitude fluctuations based on the 392 SNe~Ia at $z>0.1$ in the Union2 data set. Figure~\ref{fig:lim_union2_nx} shows the results for different values of the parameters $n_{\rm x}$ and $k_{\rm c}$ together with the observational limits established from the SN~Ia data. The amplitudes of the curves are normalised according to the values in Figure~\ref{fig:delta_all}. Note that we have zoomed in on small angular separations, since the angular covariance function increases as the angular separation decreases, and that the observational limits are based on a single data bin out to $\theta=1.5^{\circ}$ in which there are 2086 SN pairs. We point out that the observational limits are essentially unaffected by the low-redshift cut made in the data set, since the small angle bin is vastly dominated by SN pairs at higher $z$. Also plotted is the expected angular covariance from gravitational lensing based on the linear matter power spectrum calculated following Cooray~\etal~\cite{2006PhRvL..96b1301C}. The lensing curve lies below the dark energy clustering curves, but we expect it to increase at the smallest angular separations if the non-linear matter clustering contribution would be included. In fact, there are tentative detections of the gravitational lensing of SNe~Ia from non-linear matter clustering from the correlation of SN~Ia magnitude residuals and the positions and masses of foreground galaxies~\cite{2007JCAP...06..002J,2010MNRAS.402..526J}. 

The covariance induced by the dark energy fluctuations is within the observational limits and it is thus not possible to rule out any phenomenological model with linear dark energy fluctuations using the current SN~Ia data. However, given more SN data, this method can yield important constraints also for linear fluctuations. Based on Figure~\ref{fig:lim_union2_nx}, we see that we will be able to constrain linear dark energy fluctuations if the observational limits are reduced by an order of magnitude. Since the confidence limits scale roughly as $\sim 1/\sqrt{N_{\rm p}}$, we expect this to be possible if the number of SN pairs in the bin increases by a factor of 100. This corresponds to having roughly ten times more SNe~Ia with angular separations $\theta\leq 1.5^{\circ}$ compared to what is currently available.

\section{Conclusions and discussion}\label{conclusions}
Much of the current efforts in dark energy research are focussed on determining whether dark energy can be described by a cosmological constant or if it is a dynamical quantity. If we are to unveil the nature of dark energy, it is important to consider the possibility of both temporal and spatial variations. 

In this paper, we have investigated the effect of dark energy fluctuations on the angular covariance function of SN~Ia magnitude residuals. In the first part, we performed a quantitative analysis of the covariance of magnitude residuals using the Union2 data set. We showed that the data are consistent with being uncorrelated and that the covariance is constrained to $<10^{-3}$ over all angular separations. In the second part, we constructed a formalism which relates a phenomenological dark energy fluctuation power spectrum to the angular covariance of magnitude fluctuations. 
We found that the covariance induced by linear dark energy fluctuations cannot be ruled out by the current data. However, as the number of well observed SNe~Ia increases, it will be possible to discriminate between different phenomenological models. Ten times more SNe with small angle separations will suffice to rule out certain models also with small dark energy fluctuations. Larger statistics would also allow for an analysis in terms of correlations in the Hubble diagram residuals as a function of spatial distances between SNe rather than only angular separations. 

At this point, it is important to again point out the fact that the results are not exact in the sense that it is not (yet) possible to calculate cosmological distances in general inhomogeneous cosmologies. This is usually referred to as the backreaction problem, i.e., how local inhomogeneities backreact on the global expansion~\cite{2008GReGr..40..467B}. In our setup, inhomogeneities produce perturbations in the Hubble function with a fixed background expansion. In reality, these inhomogeneities might backreact, due to the non-linear structure of the field equations, to produce a different global expansion rate. In order to obtain what we believe to be at least order of magnitude correct estimates of SN magnitude angular covariance, we have assumed that local fluctuations in the dark energy density will cause local fluctuations in the Hubble expansion that in turn will induce changes in the luminosity distance to a given redshift. In doing this, we have normalised the local expansion such that $H\left(z=0\right)=H_{0}$, i.e., the Hubble constant does not have spatial variations. This is accomplished by introducing perturbations in the matter component by hand in the Hubble function such that the dark energy clustering is anti-correlated with the matter clustering. 
In a more refined formalism, one would like to include the matter and dark energy power spectrum separately. This would, however, require the inclusion of the simultaneous effect from local perturbations in the matter and dark energy component on the Hubble function. We also note that in our model, even for vanishing dark energy fluctuations, we would still expect an induced covariance from the local perturbations in the expansion history produced by the dark matter fluctuations. Thus, in a more general context, the method presented in this paper is useful for investigating Hubble flow perturbations.

Moreover, our approach is not valid in the non-linear clustering regime, and as such we cannot test phenomenological models with large density fluctuations. 
However, since we expect the problem of backreaction to be more severe for non-linear fluctuations, we have considered it superfluous to include the extra complications of allowing for large fluctuations in our simple model. The formalism also breaks down at close to horizon scales, where dark energy fluctuations could occur. In view of these limitations, we welcome all efforts to solve the problem of the expansion history and light propagation in an inhomogeneous universe. An improved formalism will definitely be warranted in the event that correlations are detected in future data.

\ack
MB acknowledges support from the HEAC Centre funded by the Swedish
Research Council. EM acknowledges support from the Swedish Research
Council.

\appendix
\section{Derivation of the weight function}\label{append}

We want to construct a weight function $W\left(r\right)$ that connects the dark energy fluctuations $\delta_{\rm x}$ with the fractional fluctuations in the luminosity distance $\delta d_{\rm L}/d_{\rm L}$. The goal is to arrive at a form
\begin{equation}\label{eq:ddlave}
\left[\frac{\delta d_{\rm L}}{d_{\rm L}}\left(\mathbf{\hat{n}}\right)\right]_{\rm average}=\intop_{0}^{r_{\rm s}}drW\left(r\right)\delta_{\rm x}\left[\mathbf{\hat{n}},z\left(r\right)\right]\ .
\end{equation}
We start with the Hubble function in some given direction $\mathbf{\hat{n}}$ (we suppress the directional dependence in the following equations) and add a perturbation in both the matter and dark energy component,
\begin{equation}\label{eq:hubblefuncpert}
H^{2}\left(z\right)=\frac{8\pi G}{3}\left[\bar{\rho}_{\rm m}\left(z\right)+\bar{\rho}_{\rm x}\left(z\right)+\delta\rho_{\rm x}\left(z\right)+\delta\rho_{\rm m}\left(z\right)\right]\ .
\end{equation}
We normalise our model such that the Hubble constant is spatially homogeneous, $H\left(z=0\right)=H_{0}$, imposing $\delta\rho_{{\rm x},0}=-\delta\rho_{{\rm m},0}$. Subscript 0 denotes values at $z=0$. Suppose that 
\begin{equation}\label{eq:drhox}
\frac{\delta\rho_{\rm x}\left(z\right)}{\bar{\rho}_{\rm x}\left(z\right)}=\frac{\delta\rho_{{\rm x},0}}{\bar{\rho}_{{\rm x},0}}\left(1+z\right)^{-s_{\rm x}/2}\ ,
\end{equation}
The parameter $s_{\rm x}$ determines the redshift dependence of the dark energy density contrast. For the matter perturbation we have
\begin{equation}\label{eq:drhom}
\frac{\delta\rho_{\rm m}\left(z\right)}{\bar{\rho}_{\rm m}\left(z\right)}=-\frac{\delta\rho_{{\rm x},0}}{\bar{\rho}_{{\rm m},0}}\left(1+z\right)^{-s_{\rm m}/2}\ .
\end{equation}
Using equation~(\ref{eq:drhox}) in equation~(\ref{eq:drhom}) then gives
\begin{equation}
\delta\rho_{\rm m}\left(z\right)=-\delta\rho_{\rm x}\left(z\right)\frac{\left(1+z\right)^{3}}{f\left(z\right)}\left(1+z\right)^{-s_{\rm m}/2+s_{\rm x}/2}\ .
\end{equation}
The perturbed Hubble function can then be written as
\begin{equation}\label{eq:hubblefuncpert2}
H^{2}\left(z\right)=\frac{8\pi G}{3}\left[\bar{\rho}_{\rm m}\left(z\right)+\bar{\rho}_{\rm x}\left(z\right)+\delta\rho_{\rm x}\left(z\right)\alpha\left(z\right)\right]\ ,
\end{equation}
where we have defined
\begin{equation}
\alpha\left(z\right)\equiv1-\frac{\left(1+z\right)^{3}}{f\left(z\right)}\left(1+z\right)^{-s_{\rm m}/2+s_{\rm x}/2}\ .
\end{equation}
Dividing equation~(\ref{eq:hubblefuncpert2}) by the critical density $\rho_{{\rm crit},0}=3H_{0}^{2}/8\pi G$ gives
\begin{equation}\label{eq:hubblefuncpert3}
H^{2}\left(z\right)=H_{0}^{2}\left[\Omega_{\rm m}\left(1+z\right)^{3}+\Omega_{\rm x}f\left(z\right)+\delta\Omega_{\rm x}\left(z\right)\alpha\left(z\right)\right]\ ,
\end{equation}
where we have defined 
\begin{equation}\label{eq:domegax}
\delta\Omega_{\rm x}\left(z\right)\equiv\frac{\delta\rho_{\rm x}\left(z\right)}{\rho_{{\rm crit},0}}=\frac{\delta\rho_{\rm x}\left(z\right)}{\bar{\rho}_{\rm x}\left(z\right)}\frac{\bar{\rho}_{\rm x}\left(z\right)}{\rho_{{\rm crit},0}}=\delta_{\rm x}\left(z\right)\Omega_{\rm x}f\left(z\right)\ .
\end{equation}
We can write the perturbed Hubble function as the unperturbed Hubble function plus a small perturbation,
\begin{equation}
H\left(z\right)=\bar{H}\left(z\right)+\delta H\left(z\right)=H_{0}E\left(z\right)+\delta H\left(z\right)\ .
\end{equation}
Since the perturbation is assumed to be small, we can Taylor expand equation~(\ref{eq:hubblefuncpert3}) to first order,
\begin{equation}
H\left(z\right)=H_{0}E\left(z\right)\left[1+\frac{\delta\Omega_{\rm x}\left(z\right)}{2E^{2}\left(z\right)}\alpha\left(z\right)\right]\ .
\end{equation}
From this we identify
\begin{equation}\label{eq:dhhbar}
\frac{\delta H\left(z\right)}{\bar{H}\left(z\right)}=\frac{\delta\Omega_{\rm x}\left(z\right)}{2E^{2}\left(z\right)}\alpha\left(z\right)\ .
\end{equation}
For a perturbation in the direction $\mathbf{\hat{n}}$, we have
\begin{equation}
d_{\rm L}+\delta d_{\rm L}\left(\mathbf{\hat{n}}\right)=c\left(1+z\right)\intop_{0}^{z}\frac{dz^{\prime}}{\bar{H}\left(z^{\prime}\right)+\delta H\left(\mathbf{\hat{n}},z^{\prime}\right)}\ .
\end{equation}
Notice that we have explicitly written the directional dependence on the perturbation in both the luminosity distance and the Hubble function, since the expansion history can be different in different directions. Doing a Taylor expansion of the integrand, we can identify
\begin{equation}
\delta d_{\rm L}\left(\mathbf{\hat{n}}\right)=-c\left(1+z\right)\intop_{0}^{z}\frac{dz^{\prime}}{\bar{H}\left(z^{\prime}\right)}\frac{\delta H\left(\mathbf{\hat{n}},z^{\prime}\right)}{\bar{H}\left(z^{\prime}\right)}\ .
\end{equation}
Shifting to the $r$ coordinate through $dr=\left[c/\bar{H}\left(z\right)\right]dz$ and using $d_{\rm L}=\left(1+z\right)r$, equation~(\ref{eq:dhhbar}) and equation~(\ref{eq:domegax}), we obtain
\begin{equation}
\frac{\delta d_{\rm L}\left(\mathbf{\hat{n}},r\right)}{d_{\rm L}\left(r\right)}=-\frac{1}{r}\intop_{0}^{r}dr^{\prime}\frac{\Omega_{\rm x}}{2E^{2}\left(r^{\prime}\right)}f\left(r^{\prime}\right)\alpha\left(r^{\prime}\right)\delta_{\rm x}\left(\mathbf{\hat{n}},r^{\prime}\right)\ .
\end{equation}
Note that, since $\alpha (z)<0$ for $z>0$, a local dark energy overdensity will increase the luminosity distance in the direction of the fluctuation. 
We can obtain the total averaged fractional fluctuation in the luminosity distance integrated over all sources in the direction $\mathbf{\hat{n}}$ using the normalised source distribution function $n(r)$,
\begin{equation}\label{eq:averddl}
\left[\frac{\delta d_{\rm L}}{d_{\rm L}}\left(\mathbf{\hat{n}}\right)\right]_{\rm average}=-\intop_{0}^{r_{\rm s}}dr\frac{n\left(r\right)}{r}\intop_{0}^{r}dr^{\prime}\frac{\Omega_{\rm x}}{2E^{2}\left(r^{\prime}\right)}f\left(r^{\prime}\right)\alpha\left(r^{\prime}\right)\delta_{\rm x}\left(\mathbf{\hat{n}},r^{\prime}\right)\ .
\end{equation}
Since the distribution function $n\left(r\right)$ will be a sum of delta functions which represents the distribution of the SNe~Ia, we can switch the integration limits. What equation~(\ref{eq:averddl}) essentially does, is to pick out each SN and look at the perturbations $\delta_{x}\left(\mathbf{\hat{n}},r\right)$ in front of it to see how they affect the luminosity distance. This will be equivalent to looking at the perturbation $\delta_{x}\left(\mathbf{\hat{n}},r\right)$ at each point $r$ and see how each SN behind it in the line of sight will be affected in terms of the luminosity distance. We can thus write equation~(\ref{eq:averddl}) as
\begin{equation}
\left[\frac{\delta d_{\rm L}}{d_{\rm L}}\left(\mathbf{\hat{n}}\right)\right]_{\rm average}=-\intop_{0}^{r_{\rm s}}dr\Biggr[\frac{\Omega_{\rm x}}{2E^{2}\left(r\right)}f\left(r\right)\alpha\left(r\right)\intop_{r}^{r_{\rm s}}dr^{\prime}\frac{n\left(r^{\prime}\right)}{r^{\prime}}\Biggl]\delta_{\rm x}\left(\mathbf{\hat{n}},r\right)\ .
\end{equation}
Comparing this expression with equation~(\ref{eq:ddlave}), we can identify the weight function $W\left(r\right)$ as
\begin{equation}
W\left(r\right)=-\frac{\Omega_{\rm x}}{2E^{2}\left[z\left(r\right)\right]}f\left[z\left(r\right)\right]\alpha\left[z\left(r\right)\right]\intop_{r}^{r_{\rm s}}dr^{\prime}\frac{n\left(r^{\prime}\right)}{r^{\prime}}\ .
\end{equation}

\section*{References}

\end{document}